\documentclass[a4paper]{article}

\usepackage{amsmath,amssymb,amsthm}
\usepackage{url}
\usepackage{xspace}
\usepackage{array}
\usepackage{microtype}

\usepackage{tikz}
\usetikzlibrary{arrows,automata,shapes,decorations.markings,decorations.pathmorphing,backgrounds,fit,snakes}
\usepgflibrary{arrows}
\tikzstyle{point}=[circle,draw]

\newcommand{\who}{Guillaume Hoffmann}
\newcommand{\what}{Undecidability of a Very Simple Modal Logic with Binding}

\author{\who\\
Consejo Nacional de Investigaciones Cient\'ificas y T\'ecnicas (CONICET)\\
Universidad Blas Pascal\\
Universidad Nacional de C\'ordoba\\
\texttt{guillaume.hoffmann@conicet.gov.ar}
}

\title{\what}
\date{2015-08-14}
\newcommand{\gM}{\mathcal{M}}

\newcommand{\down}{{\downarrow}}

\newcommand{\Hl}{\mathcal{H}}

\newcommand{\dia}[1]{\langle#1\rangle}

\newcommand{\kn}{\textcircled{k}}
\newcommand{\re}{\textcircled{r}}

\newcommand{\gB}{\mathcal{B}}

\newcommand{\pmloa}{$\gB(\dia{r},\kn)$\xspace}

\newcommand{\pmloas}{\pmloa-{\sc sat}\xspace}

\newcommand{\di}{\lozenge}

\newcommand{\bx}{\square}

\newcommand{\ra}{\rightarrow}

\makeatletter
\renewcommand{\verbatim@font}{\small\ttfamily}
\makeatother

\DeclareMathSymbol{:}{\mathbin}{operators}{"3A}

\newtheorem{definition}{Definition}

\newtheorem{lemma}{Lemma}
\newtheorem{theorem}{Theorem}

\begin{document}

\maketitle

\begin{abstract}
We show undecidability of the satisfiability problem of
what is arguably the simplest non-sub-Boolean modal
logic with an implicit notion of binding. This work
enriches the series of existing results of undecidability of
modal logics with binders, which started with Hybrid Logics and
continued with Memory Logics.
\end{abstract}


\section{Modal Logics, Names and Binders}

Modal Logics are languages that are able to describe graphs
from an internal perspective.
When they are enriched with \emph{nominals},
that is, propositional symbols that are true at only one state of the graph,
they are called \emph{Hybrid Logics} \cite{arechybr05b}.
A nominal can unequivocally name a particular state.
It was soon discovered that the ability to dynamically
name states could lead to the ability to describe more interesting graphs.
Goranko~(\cite{goranko94}) introduced the \emph{down-arrow binder}, with the syntax
$\down x . \varphi$ meaning ``after naming the
current state $x$, $\varphi$ holds''.
For instance, $\square \down x . \lozenge x$ means ``all accessible
states are reflexive''.

Hybrid Logics with binders are robustly undecidable, from the full $\Hl(@,\down)$
language~\cite{blackselig95}, to $\Hl(\down)$ with only one relation
and no propositional symbol beyond nominals~\cite{arecroad99}.
Also, $\Hl(\down)$ with propositional symbols, a single nominal and
a single relation has the same fate \cite{Marx02}.

Attempts have been made to find interesting decidable fragments of hybrid
logics with binder.
Some successful approaches involve restricting the class of models on which the satisfiability
problem is specified: models of finite width~\cite{tencate2005complexity}, or models
with a single relation and certain frame properties (S5, transitive, complete) \cite{schneider07phd}.

On the other hand, a few syntactical restrictions
have been successful.
$\Hl(@,\down) \setminus  \bx\down\bx$, the language obtained by removing formulas that contain a nesting of $\bx$, $\down$ and $\bx$, is
decidable \cite{tencate2005complexity}. Also, only allowing
one nominal, and restricting the depth between the binder and the nominal to 2
yields decidability \cite{gosc12}.

\section{Memory Logics}

\emph{Memory logics}~\cite{arechybr07,arec:expr11}
are a distinct take on the binder.
They are modal logics with the ability to \emph{store} the current state of evaluation into
some memory, and to consult whether the current state belongs to it.
The two operators associated to these
actions are $\re$ and $\kn$. $\re\varphi$ means ``remember the current state
and evaluate $\varphi$'' and $\kn$
means ``this state is known''.

Given a model $\gM = \dia{W,R,V}$, a state $s$ and a memory $S$
(which is a subset of $W$),
the semantics of these operators is given by:
$$
\begin{array}{rcl}
\gM, S, s \models \re\varphi & \mbox{ iff } & \gM, S \cup \{s\}, s\models \varphi\\
\gM, S, s \models \kn & \mbox{ iff } & s \in S.
\end{array}
$$

Let us take two pointed models that are bisimilar for the basic modal logic:

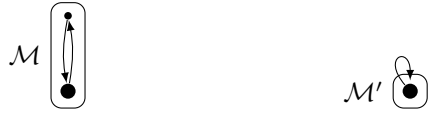
\begin{center}
\begin{tikzpicture}[>=latex]

  \node [fill=black, inner sep=2pt, circle] (l1) at (2,2) {};
  \node [fill=black, inner sep=1pt, circle] (l2) at (2,3) {};

  \path [->] (l1) edge [bend right=10] (l2) ;
  \path [->] (l2) edge [bend right=10] (l1) ;

  \node[draw,rounded corners, fit = (l1) (l2),  label=left:$\gM$ ] (b1)  {};

  \node [fill=black, inner sep=2pt, circle](r5) at (6.5,2) {} edge [in=90, out=130,loop] () ;

  \node[draw,rounded corners, fit = (r5), label=left:$\gM'$ ] (b2)  {};
\end{tikzpicture}
\end{center}

Assuming an empty initial memory, the formula $\re \lozenge \kn$ is false on the left and true on the right.
Whereas with hybrid logics we would have used an explicit name
(e.g., with the formula $\down a . \lozenge a$),
memory logics do not involve any explicit binding process.

The satisfiability problem for the basic
modal logic extended with $\re$ and $\kn$ with initial empty memory
is undecidable~\cite{arec:expr11}. Another proposal is a
memory logic without $\lozenge$ and $\re$, but with a
remember-and-move operator $\dia{r}$ equivalent to
$\re\lozenge \varphi$.
In this language, the operator $\kn$ identifies
whether the current world has already been explored during the evaluation
of the formula. It can be
seen as an ``I've already been here'' operator.
Again, satisfiability with an empty initial memory is
undecidable in this simplified language.

Undecidability proofs for the aforementioned logics regularly
involve encoding
the tiling problem into the language of satisfiable formula~\cite{Boas97theconvenience}.
Since sets of tiles are represented by propositional symbols,
we propose to study a language that lacks them.

\section{The \pmloa fragment}

We consider the language \pmloa. $\gB$ refers to the Boolean
connectors, and $\dia{r}$ and $\kn$ are the only non-Boolean logical symbols. 
For readability sake, let us write $\di$ instead of $\dia{r}$.
The syntax of \pmloa is:
$$F ~ := ~ \neg F ~ \mid ~ F_1 \wedge F_2 ~ \mid ~ \di F ~ \mid ~ \kn$$

We interpret formulas on Kripke frames $\gM = \dia{W,R}$
made of a set of states $W$ and a binary relation $R$ on $W$,
with some current evaluation state $s$ and a set $S \subseteq W$
of visited states:
$$
\begin{array}{rlcl}
\gM,S,s & \models \neg \varphi &\mbox{iff}& \gM,S,s \not \models  \varphi \\
\gM,S,s & \models \varphi \wedge \psi &\mbox{iff}& \gM, S, s \models \varphi\mbox{ and  }\gM, S, s \models \psi \\
\gM,S,s & \models \di\varphi  &\mbox{iff}&\mbox{there exists } t\in W\mbox{ such that}\\
        & & & Rst\mbox{ and } \gM,S\cup\{s\}, t \models \varphi \\
\gM,S,s & \models \kn            &\mbox{iff}&  s \in S\\
\end{array}
$$

We define the notations  $\vee$, $\rightarrow$ and $\bx$ as usual,
and $\bot \equiv \kn \wedge \neg \kn$ and $\top \equiv \kn \vee \neg \kn$.

We will use the following notion of satisfiability:
\begin{definition}
A formula $\varphi$ of \pmloa is
{\em satisfiable}, or is in \pmloas,
if there exists a model $\gM=\dia{W,R}$ and $s \in W$ such that
$\gM, \emptyset, s \models \varphi$.
\end{definition}

As a warm-up, let us see that we can define a formula
satisfiable only in infinite models. Consider the formula $Inf$
made of the following conjuncts:
\begin{quote}
\begin{enumerate}
\item $s$
\item $\bx\neg s$
\item $\bx\bx(\kn \ra s)$
\item $\di\di\kn$
\item $\bx(\di\top \rightarrow \di(\neg \kn \wedge \neg s ))$
\item $\bx\bx(\neg s \ra \di(\kn \wedge s \wedge
                        \di(\kn\wedge\bx(\kn \ra s))))$
\item $\bx\bx(\neg s \ra \bx(\neg s \ra
                   \di(\kn \wedge s \wedge \bx ( \mbox{a-or-b} \ra \di c))))$
\end{enumerate}
\end{quote}

Which use the macros:
$$
\begin{array}{lll}
s & \equiv &  \di\bx\bot\\
\mbox{a-or-b} & \equiv & \kn \wedge \di (\kn \wedge \neg s)\\
c & \equiv & \kn \wedge \bx(\kn \ra s)\\
\end{array}
$$

We call ``spy'' the evaluation state and review the conjuncts one by one,
each time assuming all previous conjuncts also hold:
\begin{enumerate}
\item spy sees a dead-end
\item spy's successors see no dead-ends, also implying spy is irreflexive
\item successors of spy are irreflexive
\item spy has a successor that has spy as a succesor
\item if a successor of spy is not a dead-end, then
      it has a successor that is different from itself
\item all states of the chain are linked to spy in both directions, and
      the relation between chain states is asymetric
\item relation between chain states is transitive
      (\mbox{a-or-b} and $c$ are macros that help referring to successive
       states of the infinite chain)
\end{enumerate}

As we require that the relation on chain states be serial (5), asymetric (6),
irreflexive (3) and transitive (7), we get an infinite chain of states
(transitive links are not represented):

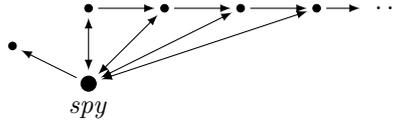
\tikzset{
    >=stealth',
    ar/.style={
           shorten <=2pt,
           shorten >=2pt,}
}

\begin{center}
\begin{tikzpicture}[>=latex]

  \node (n0)  at (0,.5) [shape=circle,draw,fill, inner sep=1pt] {} ;

  \node (n1)  at (1,1) [shape=circle,draw,fill, inner sep=1pt] {} ;
  \node (n2)  at (2,1) [shape=circle,draw,fill, inner sep=1pt] {} ;
  \node (n3)  at (3,1) [shape=circle,draw,fill, inner sep=1pt] {} ;
  \node (n4)  at (4,1) [shape=circle,draw,fill, inner sep=1pt] {} ;
  \node (n5)  at (5,1)  {\ldots} ;

  \node (spy) at (1,0) [shape=circle,draw,fill, inner sep=2pt,label=below:$\mathit{spy}$] {} ;

  \draw [ar,->] (spy) -- (n0);
  \draw [ar,<->] (spy) -- (n1);
  \draw [ar,<->] (spy) -- (n3);
  \draw [ar,<->] (spy) -- (n2);
  \draw [ar,<->] (spy) -- (n4);

  \draw [ar,->] (n1) -- (n2);
  \draw [ar,->] (n2) -- (n3);
  \draw [ar,->] (n3) -- (n4);
  \draw [ar,->] (n4) -- (n5);
 
\end{tikzpicture}
\end{center}

\begin{theorem}
If there is a frame $\gM =\dia{W,R}$ and $s\in W$ such that
$\gM,\emptyset,s \models Inf$ then $W$ is infinite.
\end{theorem}

Hence \pmloa lacks the finite model property.
This does not necessarily imply that this logic is undecidable,
but we are going to see that it is actually the case.

\section{Undecidability of \pmloas}

We adapt the proof of \cite{Marx02}, where the tiling problem~\cite{Boas97theconvenience}
is encoded into a description logic with the down-arrow binder, one
nominal, and one relation.

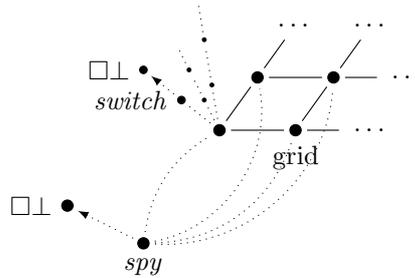
\begin{figure}[h!]
\begin{center}
\begin{tikzpicture}[>=latex]

\node (g11) at (2,2) [shape=circle,draw, fill, inner sep=1.5pt] {} ;
\node (g12) at (3,2) [shape=circle,draw,fill, inner sep=1.5pt,label=below:grid] {} ;
\node (g13) at (4,2) [] {\ldots} ;
\node (g21) at (2.5,2.7) [shape=circle,draw,fill, inner sep=1.5pt] {} ;
\node (g22) at (3.5,2.7) [shape=circle,draw,fill, inner sep=1.5pt] {} ;
\node (g23) at (4.5,2.7) [] {\ldots} ;
\node (g31) at (3,3.4) [] {\ldots} ;
\node (g32) at (4,3.4) [] {\ldots} ;

\draw [ar,-] (g11) -- (g12);
\draw [ar,-] (g12) -- (g13);
\draw [ar,-] (g21) -- (g22);
\draw [ar,-] (g22) -- (g23);

\draw [ar,-] (g11) -- (g21);
\draw [ar,-] (g21) -- (g31);
\draw [ar,-] (g22) -- (g32);
\draw [ar,-] (g12) -- (g22);

\node (spy) at (1,0.5) [shape=circle,draw,fill,inner sep=1.5pt,label=below:$\mathit{spy}$] {} ;
\node (blind) at (0,1) [shape=circle,draw,fill,inner sep=1.5pt, label=left:$\bx\bot$] {} ;

\draw [ar,->, dotted] (spy) -- (blind);
\path [ar,-,dotted] (spy) edge [bend left=25] (g11);
\path [ar,-,dotted] (spy) edge [bend right=45] (g21);
\path [ar,-,dotted] (spy) edge [bend right=35] (g12);
\path [ar,-,dotted] (spy) edge [bend right=45] (g22);

\node (p00) at (1.5,2.4) [shape=circle,draw,fill, inner sep=1pt,label=left:$\mathit{switch}$] {} ;
\node (p01) at (1,2.8) [shape=circle,draw,fill, inner sep=1pt, label=left:$\bx\bot$] {} ;

\node (p10) at (1.8,2.4) [shape=circle,draw,fill, inner sep=0.5pt] {} ;
\node (p11) at (1.6,2.8) [shape=circle,draw,fill, inner sep=0.5pt] {} ;
\node (p12) at (1.4,3.2) [inner sep=0.5pt] {} ;

\node (p20) at (1.9,2.6) [shape=circle,draw,fill, inner sep=0.5pt] {} ;
\node (p21) at (1.8,3.2) [shape=circle,draw,fill, inner sep=0.5pt] {} ;
\node (p22) at (1.7,3.8) [inner sep=0.5pt] {} ;

\draw [ar,-, dotted] (g11) -- (p00);
\draw [ar,->, dotted] (p00) -- (p01);

\draw [ar,dotted] (g11) -- (p10);
\draw [ar,dotted] (p10) -- (p11);
\draw [ar,dotted] (p11) -- (p12);

\draw [ar,dotted] (g11) -- (p20);
\draw [ar,dotted] (p20) -- (p21);
\draw [ar,dotted] (p21) -- (p22);

\end{tikzpicture}
\end{center}
\caption{Model of a formula encoding some tiling problem,
         with details for the first grid state}
\label{pmlo_grid_model}
\end{figure}

We build a formula that is satisfiable only in models with the following
components (see Figure~\ref{pmlo_grid_model}):
\begin{itemize}
\item the \emph{spy state} is where the formula is evaluated.
\item \emph{grid states} are direct successors of the spy state.
A grid state is connected to other grid states, but also to
switch and propositional states.
\item \emph{switch states} are in one-to-one correspondence with grid states,
and help building the structure by simulating an explicit binding operator.
A switch state sees the grid state that sees it.
\item \emph{propositional states} encode propositional symbols
      at every grid state, using chains of determined length.
      A grid state can see several propositional states, but a
      propositional state does not see its origin grid state.
\item The spy, switch and propositional states see at least one dead-end state,
      while grid states do not.
\end{itemize}

Let us enumerate the conjuncts of the formula that
encodes a given instance of the tiling problem.

The first formulas specify that the spy state
sees a dead-end, is irreflexive, that all its successors
are also irreflexive, that all grid states see
the spy state, and that the relation between grid states is
symmetric:
\begin{enumerate}
\item $s$
\item $\bx \neg s$
\item $\bx\bx(\kn ~ \ra ~ s)$
\item $\bx( \di \top ~  \ra ~ \di \kn )$
\item $\bx\bx ( \neg s \ra  \di ( \kn \wedge \neg s))$
\end{enumerate}

The following two formulas implement switch states for
grid states accessible in one and two steps from spy:
\begin{enumerate}
\setcounter{enumi}{5}
\item $\bx( \di\top \ra  \di sw)$
\item $\bx\bx( \neg s \ra  \di sw)$
\end{enumerate}

With the macro:
$$
\begin{array}{lll}
sw & \equiv & \neg\kn \wedge s \\
   & \wedge & \di(\kn \wedge \neg s)\\
   & \wedge & \bx ( \di\top ~ \ra (\kn \wedge \neg s \wedge \bx( (s \wedge \di\kn)  \ra \kn)))
\end{array}
$$

Each grid state sees to its own switch state,
which sees a dead-end and sees the grid state that sees it.
Switch states help us simulate the $\re$ operator:
to ``remember'' a particular grid state, we visit its
associated switch state and come back. Note that formula $(7)$
depends on formula $(6)$ to implement switches on grid states
distants from two steps from spy.

We use the following macro to ``remember' the current grid state:
$$
\begin{array}{lll}
\re'\varphi  & \equiv & \bx(\neg\kn \wedge s  ~ \ra \bx(\kn  ~ \ra \varphi ))\\
\end{array}
$$

And to check that a grid state has been remembered:
$$
\begin{array}{lll}
\kn' & \equiv & \kn \wedge \bx((s \wedge \di\kn) \ra \kn)\\
\end{array}
$$

Now the following formula makes the spy state see
every grid state:

\begin{enumerate}
\setcounter{enumi}{7}
\item $\bx\bx( \neg s \ra \re'\bx\bx( \kn \wedge s ~ \ra \di \kn'))$
\end{enumerate}

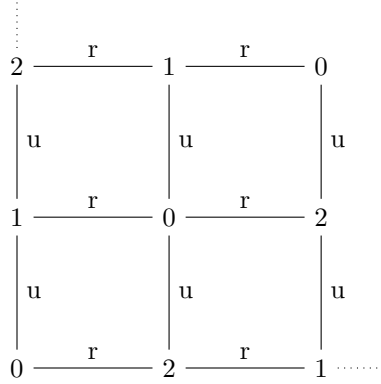
\begin{figure}[h!]
\begin{center}
\begin{tikzpicture}[>=latex]

\node (g00) at (0,0) [inner sep=1.5pt] {0} ;
\node (g10) at (2,0) [inner sep=1.5pt] {2} ;
\node (g20) at (4,0) [inner sep=1.5pt] {1} ;
\node (g01) at (0,2) [inner sep=1.5pt] {1} ;
\node (g11) at (2,2) [inner sep=1.5pt] {0} ;
\node (g21) at (4,2) [inner sep=1.5pt] {2} ;
\node (g02) at (0,4) [inner sep=1.5pt] {2} ;
\node (g12) at (2,4) [inner sep=1.5pt] {1} ;
\node (g22) at (4,4) [inner sep=1.5pt] {0} ;

\node (g30) at (5,0) [inner sep=1.5pt] {} ;
\node (g03) at (0,5) [inner sep=1.5pt] {} ;

\draw [ar,-] (g00) -- node [label=r] {} (g10);
\draw [ar,-] (g10) -- node [label=r] {} (g20);
\draw [ar,-] (g01) -- node [label=r] {} (g11);
\draw [ar,-] (g11) -- node [label=r] {} (g21);
\draw [ar,-] (g02) -- node [label=r] {} (g12);
\draw [ar,-] (g12) -- node [label=r] {} (g22);

\draw [ar,-] (g00) -- node [label=right:u] {} (g01);
\draw [ar,-] (g01) -- node [label=right:u] {} (g02);
\draw [ar,-] (g10) -- node [label=right:u] {} (g11);
\draw [ar,-] (g11) -- node [label=right:u] {} (g12);
\draw [ar,-] (g20) -- node [label=right:u] {} (g21);
\draw [ar,-] (g21) -- node [label=right:u] {} (g22);

\draw [ar,-,dotted] (g20) -- (g30);
\draw [ar,-,dotted] (g02) -- (g03);

\end{tikzpicture}
\end{center}
\caption{Organization of the grid}
\label{grid}
\end{figure}

We encode the tiling problem with only one relation,
by combining a symmetric relation with ``propositional macros''
to represent two
``meta-relations'' \emph{up} and \emph{right},
as shown in Figure~\ref{grid}.
In what follows, we use the notations $i$, $s(i)$ and $p(i)$, with:
$$
\begin{array}{lll}
i & \in & \{0,1,2\}\\
s(i) & = & (i + 1)\mbox{ mod }3\\
p(i) & = & (i + 2)\mbox{ mod }3\\
\end{array}
$$

Every grid state has an ``up''- and a ``right''-successor:

\begin{enumerate}
\setcounter{enumi}{8}
\item
 $\bigwedge_{i \in \{0,1,2\}}
 \bx(\di\top \wedge i \ra \di(\neg s \wedge u \wedge \di( \neg s \wedge s(i) )))$
\item
 $\bigwedge_{i \in \{0,1,2\}}
 \bx(\di\top \wedge i \ra \di(\neg s \wedge r \wedge \di( \neg s \wedge p(i) )))$
\end{enumerate}

``Up'' and ``right'' are irreflexive and functional:
\begin{enumerate}
\setcounter{enumi}{10}
\item
 $\bigwedge_{i \in \{0,1,2\}}
 \bx( \di\top \wedge s(i) \ra \re'\bx(\neg s \wedge u \ra \bx( \neg s \wedge i \ra \neg \kn \wedge \bx( \neg s \wedge u \ra \bx( \neg s \wedge s(i) \ra \kn' )))))$
\item 
 $\bigwedge_{i \in \{0,1,2\}}
 \bx( \di\top \wedge p(i) \ra \re'\bx(\neg s \wedge r \ra \bx( \neg s \wedge i \ra \neg \kn \wedge \bx( \neg s \wedge r \ra \bx( \neg s \wedge p(i) \ra \kn' )))))$
\end{enumerate}

\begin{figure}[h!]
\begin{center}
\begin{tikzpicture}[>=latex]

\node (g00) at (0,0) [inner sep=1.5pt] {$i$} ;
\node (g10) at (2,0) [inner sep=1.5pt] {$p(i)$} ;
\node (g01) at (0,2) [inner sep=1.5pt] {$s(i)$} ;
\node (g11) at (2,2) [inner sep=1.5pt] {$i$} ;

\draw [ar,-] (g00) -- node [label=r] {} (g10);
\draw [ar,-,dotted] (g01) -- node [label=r] {} (g11);

\draw [ar,-] (g00) -- node [label=right:u] {} (g01);
\draw [ar,-,dotted] (g10) -- node [label=right:u] {} (g11);

\end{tikzpicture}
\end{center}
\caption{Confluence property}
\label{confluence}
\end{figure}
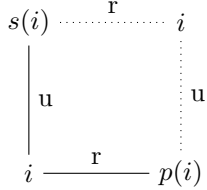

We finish with the confluence property as shown in Figure~\ref{confluence}:
\begin{enumerate}
\setcounter{enumi}{12}
\item
 $\bigwedge_{i \in \{0,1,2\}}
 \bx(\di\top \wedge s(i) \ra \re'\bx(\neg s \wedge u \ra \bx (\neg s \wedge i \ra \bx( \neg s \wedge r \ra \bx ( \neg s \wedge p(i) \ra \di ( \neg s \wedge u \wedge \di (  \neg s \wedge i \wedge \di( \neg s \wedge r \wedge \di( \neg s \wedge s(i) \wedge \kn')))))))))$
\end{enumerate}

Now, we encode the actual tiling problem.
Given $T = {T_1, \ldots, T_k}$ a set of tile types, we represent
each type $T_i$ as a macro $t_i$.
At every point of the grid, one and only one tile holds:
\begin{enumerate}
\setcounter{enumi}{13}
\item
 $\bigwedge_{i \in \{0,1,2\}}
 \bx( \di\top \wedge i  \ra (\bigvee_{1 \leq n \leq k}  t_n)  \wedge (\bigwedge_{1 \leq n < m \leq k} \neg (t_n \wedge t_m)) )$
\end{enumerate}

Then, tile matching:
\begin{enumerate}
\setcounter{enumi}{14}
\item
 $\bigwedge_{i \in \{0,1,2\}, 1 \leq n \leq k}
 \bx( \di\top \wedge i \wedge t_n \ra \square ( \neg s \wedge u \ra \square ( \neg s \wedge s(i) \ra \bigvee
 \{ t_m \mid top(T_n) = bottom(T_m) \} ))  )$
\item
 $\bigwedge_{i \in \{0,1,2\}, 1 \leq n  \leq k}
 \bx( \di\top  \wedge i \wedge t_n \ra \square ( \neg s \wedge r \ra \square ( \neg s \wedge p(i) \ra \bigvee
 \{ t_m \mid right(T_n) = left(T_m) \} )) )$
\end{enumerate}

We initialize the grid with a first state:
\begin{enumerate}
\setcounter{enumi}{16}
\item $\di(\di\top \wedge 0)$
\end{enumerate}

Finally,
we assign a distinct natural number $p$ to every propositional macro
that appear in formulas 9--17, and rewrite it to:
\vspace{-0.4cm}
$$
\begin{array}{lll}
p & \equiv & \di(s \wedge \neg\kn \wedge \bx\neg\kn \wedge  \overbrace{\di\di\ldots\di}^{p ~ times}\bx\bot )
\end{array}
$$

We call ``propositional states'' the states accessible from
grid states where $s \wedge \neg \kn \wedge \bx\neg \kn$ holds.
Hence, propositional states do not see the grid point that see them.
Each ``propositional symbol'' that should be true at a grid state
is represented by a propositional state that sees a dead-end, does not see
the grid point where it came from, and sees the beginning
of a chain of determined length.
Propositional states and their corresponding chains are
not seen by the spy state (see formula $(8)$).
One can check that negating such a macro does not
interfere with the structure of the grid:
$$
\begin{array}{lll}
\neg p & \equiv & \bx(s \wedge \neg\kn \wedge \bx\neg\kn  ~ \ra  \overbrace{\di\di\ldots\di}^{p ~ times}\bx\bot )
\end{array}
$$

\begin{lemma}
Given a tiling problem $T = T_1 \ldots T_k$, let
$Grid(T)$ be the conjunction of formulas 1--17
rewritten with the propositional encoding previously shown.

$T$ tiles the grid if, and only if, $Grid(T) \in$ \pmloas.
\end{lemma}

\begin{theorem}
\pmloas is undecidable.
\end{theorem}

\section{Closing Remarks}

We showed that \pmloa, arguably the simplest non-sub-Boolean modal logic with binding,
has an undecidable satisfiability problem.
This result can be reused to show undecidability of the satisfiability problem of most
memory logics and hybrid logics with binders, by means of simple
satisfiability-preserving translations. For instance the following
translation from \pmloas to $\Hl(\down)$-{\sc sat} \cite{arec:expr11}:
$$
\begin{array}{lll}
{\sf Tr}_N(\neg \varphi) & = & \neg {\sf Tr}_N(\varphi)\\
{\sf Tr}_N(\varphi \wedge \psi) & = & {\sf Tr}_N(\varphi) \wedge {\sf Tr}_N(\psi)\\
{\sf Tr}_N(\di\varphi) & = & \down i . {\sf Tr}_{N\cup\{i\}}(\varphi) ~ \mbox{with $i\notin N$}\\
{\sf Tr}_N(\kn) & = & \bigvee_{i\in N} i
\end{array}
$$
where $N$ is a subset of the set of nominals, initially empty.

Taking as a starting point the
basic hybrid logic with binder $\Hl(@,\down)$,
the undecidability results for modal logics with binding have
progressed in two directions. First, by maintaining the ability to bind an arbitrary number
of states, while decreasing the expressive power of the language.
This is how we went from $\Hl(@,\down)$, to $\Hl(\down)$ without
non-nominal propositional symbols \cite{arecroad99}, to memory logics with $\re$
and $\kn$  \cite{arec:expr11}, to memory logics with $\dia{r}$ instead of $\re$ and $\di$,
and finally, to the memory logic presented in this article.

The second direction has been to restrict the number of bindable
states to only one \cite{Marx02}. This approach actually
removes the differences between hybrid and memory logics
(the latter of which were not yet invented at the time of the result).

It is not clear how to transpose this last approach to the logic presented
here. One way would be to restrict the memory to the last $n$
visited states.
But this involves giving
up on the simplicity of the very definition of the problem.
Hence it seems probable
that the quest for pushing the undecidable frontier will continue
within a different setting.

\bibliographystyle{plain}
\bibliography{memory}

\end{document}